\documentclass[10pt, conference, letterpaper]{IEEEtran}
\usepackage{balance}
\usepackage{comment}
\usepackage{pdfpages} 
\usepackage{amsmath}
\usepackage{epstopdf}
\usepackage{amsmath}
\usepackage{cite}
\usepackage{url}

\usepackage{color}

\usepackage{algorithm}
\usepackage{algorithmic}
\usepackage{subfigure}
\usepackage{amssymb, amsmath,graphicx,charter, latexsym}
\usepackage{enumerate}

\newtheorem{definition}{Definition}
\newtheorem{lemma}{Lemma}

\newtheorem{theorem}{Theorem}

\newcommand{\argmin}{\arg\!\min}
\newcommand{\argmax}{\arg\!\max}
\newcommand{\floor}[1]{\lfloor #1 \rfloor}
\begin{document}

\title{On the Capacity Requirement for Arbitrary End-to-End Deadline and Reliability Guarantees in Multi-hop Networks}

%\numberofauthors{2}
%\author{
%   \alignauthor Han Deng \\
%     \affaddr{Department of ECE\\
%     Texas A\&M University}\\
%     \affaddr{College Station, TX 77840, USA}
%    \email{hdeng@email.tamu.edu}
%   \alignauthor I-Hong Hou \\
%    \affaddr{Department of ECE\\
%    Texas A\&M University}\\
%    \affaddr{College Station, TX 77840, USA}
%    \email{ihou@tamu.edu}
%}
%
%
%\makeatletter
%\def\@copyrightspace{\relax}
%\makeatother

\author{\IEEEauthorblockN{Han Deng}
\IEEEauthorblockA{Department of ECE\\
Texas A\&M University\\
College Station, TX 77840, USA\\
Email: hdeng@email.tamu.edu}
\and
\IEEEauthorblockN{I-Hong Hou}
\IEEEauthorblockA{Department of ECE\\
Texas A\&M University\\
College Station, TX 77840, USA\\
Email: ihou@tamu.edu}
}

\maketitle

\begin{abstract}
It has been shown that it is impossible to achieve both stringent end-to-end deadline and reliability guarantees in a large network without having complete information of all future packet arrivals. In order to maintain desirable performance in the presence of uncertainty of future packet arrivals, common practice is to add redundancy by increasing link capacities. This paper studies the amount of capacity needed to provide stringent performance guarantees. We propose a low-complexity online algorithm and prove that it only requires a small amount of redundancy to guarantee both end-to-end deadline and reliability. Further, we show that in large networks with very high reliability requirements, the redundancy needed by our policy is at most twice as large as a theoretical lower bound. Also, for practical implementation, we propose a fully distributed protocol based on the previous centralized policy. Without adding redundancy, we further propose a low-complexity order-optimal online policy for the network. Simulation results also show that our policy achieves much better performance than other state-of-the-art policies.
\end{abstract}

%
% End generated code
%

%
%  Use this command to print the description
%
%\printccsdesc

% We no longer use \terms command
%\terms{Theory}
%
%\keywords{Multi-hop Network; Online Scheduling; CompetitiveRatio; Capacity Performance Trade-off}

\section{Introduction}
\label{section:intro}

Many emerging safety-critical applications, such as Internet of Things (IoT) and Cyber-Physical Systems (CPS), require communication protocols that support strict end-to-end delay and reliability guarantees for all packets. In a typical scenario, when sensors detect unusual events that can cause system instability, they send out this information to actuators or control centers. This information needs to be delivered within a strict deadline for actuators or control centers to resolve the unusual events. The system can suffer from a critical fault when a small portion of packets fail to be delivered on time.

Despite the huge literature on quality of service (QoS), there is little work that can provide end-to-end delay and reliability guarantees simultaneously, especially when packet arrivals are time-varying and unpredictable. The lack of progress is mainly caused by two fundamental challenges. On one hand, it is obvious that one cannot design the optimal network policies without obtaining complete knowledge of future packet arrivals and incurring high computation complexity. Therefore, practical solutions need to rely on online suboptimal policies. On the other hand, in a multi-hop network, the scheduling decision of one communication link will impact the decisions of subsequent links. The negative effects of suboptimal decisions by online policies therefore get accumulated along the path of multi-hop transmissions. In fact, a recent work by Mao, Koksal, and Shroff \cite{multihoponline} has proved that the performance of any online policies deteriorates as the length of the longest path in the network increases. As a result, no online policy can provide meaningful performance guarantees when the size of the network is large.

In order to maintain desirable performance using online suboptimal policies, current practice is to add redundancy into the system. During system deployment, the capacities of communication links are chosen to be larger than necessary. Such redundancy alleviates the negative impacts of suboptimal decisions by online policies. Using this approach, a critical question is to determine the amount of redundancy needed to provide the desirable performance guarantees. This paper aims to answer this question.

We first show that the problem of maximizing the number of timely packet deliveries can be formulated as a linear programming problem when one knows the complete knowledge of all future packet arrivals. In the setting of online policies, some of the parameters of this linear programming problem will only be revealed when the corresponding packets arrive. Therefore, online policies need to make routing and scheduling decisions for packets without knowing all parameters. On the other hand, we also observe that adding redundancy by increasing link capacities is equivalent to relaxing a subset of constraints in the linear programming problem. Based on these observations, we define a competitive ratio that, given the amount of redundancy, quantifies the relative performance of online policies in comparison to the optimal offline solution.

Using the primal-dual method, we propose an online policy that achieves good performance in terms of competitive ratio. This policy has several important features: First, when there is no redundancy added to the system, the performance of our online policy is asymptotically better than that of the recent work \cite{multihoponline} when the size of the network increases. Second, we also show that only a small amount of redundancy is needed to achieve strict performance guarantees. Specifically, in order to guarantee the timely delivery of at least $1-\frac{1}{\theta}$ as many packets as the optimal solution in a network whose longest path has length $L$, our policy only needs to increase link capacities by $\ln L+\ln \theta$ times. Finally, we also show that our policy can be implemented with very low complexity.

Next, we establish a theoretical lower bound of competitive ratio for all online policies. We show that, in order to guarantee a certain degree of performance, the redundancy needed by our policy is only a small amount away from the theoretical limit. In particular, when both $L$ and $\theta$, as defined in the previous paragraph, go to infinity, the redundancy needed by our policy is at most twice as large as the theoretical limit.

We also study online policies when one cannot increase network capacity by adding redundancy. We propose another online policy and prove that it is order optimal with fixed link capacity. Specifically, we show that this online policy guarantees to deliver at least $\frac{1}{O(\log_L)}$ as many packets before their deadlines as the optimal offline solution, where $L$ is the maximum route length. As the previous study \cite{multihoponline} has proved no online policy can deliver more than $\frac{1}{O(\log_L)}$ packets without redundancy, our policy is order-optimal.

While neither of our online policies need any information about future packet arrivals to make routing and scheduling decisions, they are centralized algorithms that require tight coordination. For large networks without a centralized coordinator, we also propose a fully distributed protocol that is inspired by the design principles of our centralized online policies. This distributed protocol only requires each node to broadcast its local congestion information very infrequently, and therefore it only incurs a small amount of communication overhead. When a packet arrives at a source node, the source node determines a suggested route for the packet using its received congestion information, and each link on the route makes scheduling decisions solely based on its local information.% Both 
%Since it is not practical to implement centralized algorithm on our current networks, we propose a heuristic distributed implementation of our centralized online algorithm. We assume the network information on all links is broadcasted periodically. Before the next broadcasting, the source node will use the previous received network information to suggest a routing and scheduling scheme for the packets. However, during the packet transmitting, each node still has certain freedom to choose when to transmit the packet on the suggested link based on the node's real-time link information.

All three of our policies are evaluated by simulations. We compare our policies with the widely used earliest deadline first policy (EDF) and recent policy studied in \cite{multihoponline}. Simulation results show that all our policies perform better than the other two policies. This result is in particular surprising because our distributed protocol even achieves better performance than the online policy in \cite{multihoponline}, which is a centralized one.%The results show that two policies both outperform EDF and recent work in \cite{multihoponline}. Also, results further verify that our order-optimal algorithm achieves better performance than our other policy.

The rest of the paper is organized as follows.
Section \ref{section:related} reviews some existing works. 
Section \ref{section:model} introduces our system model defines the competitive ratio. 
Section \ref{section:algorithm} proposes our online policy and studies its competitive ratio and computation complexity.
Section \ref{section:lower bound} establishes a theoretical lower bound of competitive ratio.
Section \ref{section:modified alg} proposes an order-optimal policy and studies its competitive ratio.
Section \ref{section:distributed} proposes a distributed protocol based on the intuitions of our centralized online policy.
Section \ref{section:simulation} provides simulation on our proposed algorithms and compare them with two other online policies. 
Finally, Section \ref{section:conclusion} concludes this paper.

\section{Related Work}
\label{section:related}
Online scheduling problem in real-time environment has been studied in many previous works. Studies show that earliest deadline first algorithm (EDF)  \cite{EDF, EDFLLF} and least laxity first algorithm (LLF) \cite{EDFLLF} achieve the same performance as the optimal offline algorithm when the system is under-loaded, that is, the optimal offline algorithm can serve all jobs in the system. In under-loaded system, all jobs enter the system can be served by EDF and we do not need to drop any job when it arrives at the system. However, in over-loaded system, even with optimal offline algorithm, there are still some jobs that cannot be served. 
%Since some jobs will be dropped, the is an additional problem of admission control. 
EDF and LLF achieve the same performance as the optimal online policy when the system is over-loaded. Also \cite{EDFLLF} proved that no online algorithm can guarantee to serve more than $1/4$ of the jobs that can be served by optimal offline algorithm and provided an algorithm in a uniprocessor system which achieves $1/4$ service bound. \cite{AC1, AC2} consider admission control in online scheduling. In \cite{AC1}, when all jobs have equal length, the competitive ratio of deterministic algorithm is bounded by 2.  \cite{AC3} considers the similar model as \cite{AC1}. It introduces a parameter $k$ to indicate the willingness of a job to have a delay before being served. It shows that when all jobs have equal length, the competitive ratio of deterministic algorithm is $(1+1/(\floor{k}+1))$-competitive instead.

In addition, online scheduling with multiple-server case has also been studied. \cite{MultiP1} studies the scheduling of equal length jobs on two identical machines. \cite{ MultiP3, MultiP4, MultiP2} studies the case with parallel machines. The scheduler need to decide whether to accept or reject a packet and which machines is chosen to serve the job. \cite{MultiP4} has proposed an algorithm with immediate decision which approaches $\frac{e}{e-1}$-competitive when the number of machines is greater or equal to 3. It also provide another lower bound that deterministic online algorithm with immediate decision is no better than 1.8-competitive when there are 2 machines. Later \cite{MultiP2} has shown that online algorithm which makes immediate decision upon job releasing is bounded by $\frac{e}{e-1}$-competitive for multiple machine case. 

There are also many works studying the scheduling problem in multihop network. 
An early study \cite{MultiH1} focuses on the problem of packet scheduling with arbitrary end-to-end delay, fix route, and known packet injection rate. It propose a distributed algorithm which achieve a certain delay bound. \cite{MultiH2} studies the scheduling problem on a tree network. Packets arrive at an arbitrary node and they need to be transmitted to root node before the deadlines. Any packet that cannot arrive root node within deadline is considered lost. Thus this is also a fix route problem. The goal is to minimize the total lost packets. Shortest time to extinction (STE) algorithm is proposed and it is shown to achieve the performance of optimal offline policy. 
Also there are many works studying the end-to-end delay in multiohop network. 
Rodoplu {\em et~al.} \cite{Rodoplu} have studied
the problem of dynamic estimating end-to-end delay over multi-hop mobile wireless networks.
Sanada Komuro and Sekiya\cite{Sanada} have used Markov-chain model to study the string-topology multi-hop network and analyse the end-to-end throughput and delay.
Jiao {\em et~al.} \cite{e2edelayanalysis} have studied the problem of estimating the end-to-end delay distribution for general traffic arrival model and Nakagami-m channel model by analyzing packet delay at each hop. Li {\em et~al.} \cite{mine2edelay} ] have proposed using expected end-to-end delay for selecting path in wireless mesh networks. The expected end-to-end delay takes both queuing delay and delay caused by unsuccessful wireless transmissions. However, their work only aims at minimizing the average end-to-end delays, and cannot provide guarantees on per-packet delays.
 
Li and Eryilmaz \cite{e2e1} has studied the end-to-end deadline constrained traffic scheduling in multihop network. They develop algorithms to meet the deadline and throughput requirement in a wired network. However, they only consider the fix route model and they do not provide any performance guarantee. 
Wang {\em et~al.} \cite{e2eroutwsn} have studied the problem of routing and scheduling on multi-hop wireless sensor network in order to optimize the system with the constraint of end-to-end delay and proposed a sub-optimal algorithm. Hou\cite{hou} proposed a throughput optimal policy for up-link tree  networks with end-to-end delay constraints and delivery ratio requierement. The packets deadlines are the end of the  frames in which they are generated. Singh and Kumar\cite{singh} have proposed a scheduling policy which maximize the throughput for multi-hop wireless networks. However, the paper uses a fix-route model and does not consider end-to-end delay. Mao, Koksal and Shroff \cite{multihoponline} also considers a fix route problem. The network has arbitrary packet arrival and packet weight. The paper aims to maximize the total cumulative weight of packets that reach destination before their deadline. The paper has proved that the competitive ratio of any online policy is no better than $O(\log{L})$, where $L$ is the length of the maximum route. It has also proposed an admission control and packet scheduling policy and shown that it is $O(L\log{L})$-competitive. Liu and Yang \cite{e2erouting} have studied the multi-hop routing problem with hard end-to-end delay and the throughput region. They also assume that packets are required to be delivered to destination within one frame and the performance is evaluated by simulation. Our work will focus on online routing and scheduling on multi-hop network with end-to-end delay constraint and aim to guarantee both packet deadline and network delivery ratio.

\section{System Model}
\label{section:model}
We consider a network with multihop transmissions. The network is represented by a directed graph where each node represents a router and an edge from one node to another represents a link between the corresponding routers. Packets arrive at their respective source nodes following some unknown sequence. We use $\mathcal{M}$ to denote the set of all packets and $\mathcal{L}$ the set of all links. When a packet $m\in\mathcal{M}$ arrives at its source node, it specifies its destination and a deadline. The packet requests to be delivered to its destination before its specified deadline. Packets that are not delivered on time do not have any value, and can be dropped from the network. We aim to deliver as many packets on time as possible.

We assume that time is slotted and numbered by $t=\{1,2,3,\dots\}$. Different links in the network may have different link capacities, and we denote by $C_l$ the number of packets that link $l$ can transmit in a time slot. At the beginning of each time slot, each node decides which packets to transmit over its links, subject to capacity constraints of the links. Packets transmitted toward a node in time slot $t$ will be received by that node at the end of the time slot, so that the node can transmit these packets to subsequent nodes starting from time slot $t+1$.

\begin{figure}
	\begin{center}
	\includegraphics[width=2.8in]{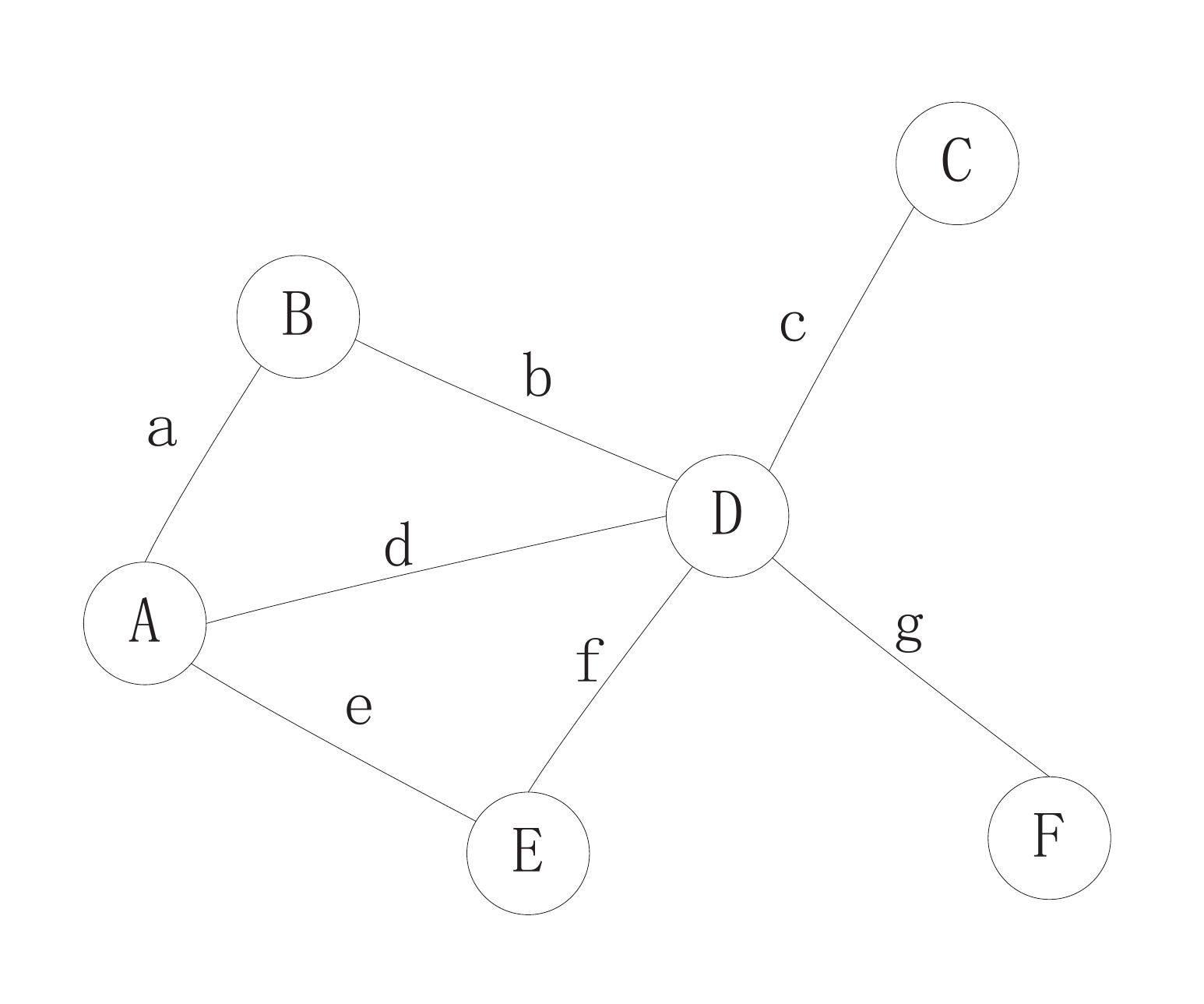}
	\caption{Network topology.}
	\label{fig:network}
	\end{center}
\end{figure}

Delivering a packet to its destination before its deadline require determining two things: the route used to forward the packet from its source to its destination, and the times at which the packet is transmitted along its route. We define a \emph{valid schedule} for each packet $m$ as the collection of links of a route, as well as the times of transmissions for each of these links, so that packet $m$ can be delivered to its destination on time. For example, consider the network shown in Fig. \ref{fig:network}. Suppose a packet arrives at node A at time slot 1, and needs to be delivered to node F before the end of time slot 3. One valid schedule for this packet is to transmit it over link d in time slot 1, and then over link g in time slot 2. We use $\{(d,1), (g,2)\}$ to represent this valid schedule. Other valid schedules include $\{(d,1), (g,3)\}, \{(e,1), (f,2), (g,3)\}$, etc. On the other hand, $\{(d,1), (g,4)\}$ is not a valid schedule because the packet is delivered to its destination after its deadline at time slot 4. The schedule $\{(d,3), (g,2)\}$ is not valid because it would require node D to transmit the packet over link g at time slot 2 before it receives the packet at time slot 3. For each packet $m$, we let $V(m)$ denote the set of valid schedules for $m$. The problem of deciding how to deliver packets on time then becomes one of choosing valid schedules for packets.

We use $X_{mk}$ to denote the schedule selection for packet $m$. If $X_{mk}=1$, packet $m$ is transmitted using valid schedule $k$, and $X_{mk}=0$, otherwise. Given the information of all packets, the problem of maximizing the total number of successful deliveries can be formulated as the following linear programming problem:

\textbf{Schedule:}
\begin{align}
Max &\sum_{m,k:k\in V(m)}X_{mk} \label{Schedule1} \\
s.t.&\sum_{k:k\in V(m)}X_{mk}\leq 1, \forall m\in \mathcal{M},\label{Schedule2}\\
	&\sum_{m,k:(l,t)\in k} X_{mk} \leq C_l, \forall l\in\mathcal{L},t\in\{1,2,\dots\},  \label{Schedule3}\\
	& X_{mk}\geq 0, \forall  m\in \mathcal{M},k\in V(m). \label{Schedule4}
\end{align}

Since $X_{mk}=1$ if packet $m$ is transmitted using valid schedule $k$, Eq. \eqref{Schedule1} is the total number of packets that are delivered on time. Eq. \eqref{Schedule2} states that at most one valid schedule can be chosen for each packet. Eq. \eqref{Schedule3} states that each link can transmit at most $C_l$ packets in any time slot. In practice, $X_{mk}$ can only be either 0 or 1, but our problem formulation allows $X_{mk}$ to be any real number in $[0,1]$. Thus, the optimal solution to \textbf{Schedule} describes an upper bound on the total number of successful deliveries.

If information of all packets is available when the system starts, the optimal solution to \textbf{Schedule} can be found by standard linear programming methods. In practice, however, packets arrive sequentially, and we need to rely on  online policies that determines the values of $X_{mk}$ for each arriving packet $m$ without knowing future packet arrivals. Without the knowledge of future arrivals, it is obvious that online policies cannot always achieve the optimal solution to \textbf{Schedule}. In fact, a recent work \cite{multihoponline} has shown that, when the longest path between a source node and a destination node is $L$, no online policy can guarantee to deliver more than $\frac{1}{\log_2 L}$ as many packets as the optimal solution. To put this number in perspective, consider a medium-sized network with $L=8$. Even when the optimal solution can deliver all packets on time, the bound in the recent work states that no online policy can guarantee to deliver more than $\frac{1}{\log_2 8}= \frac{1}{3}$ of all packets. Such performance of online policies is unacceptable for virtually any applications.

In order to achieve good performance for online policies in the presence of unknown future arrivals, we consider the scenario where service providers can increase link capacities by, for example, upgrading network infrastructures. When the link capacities are increased by $R$ times, link $l$ can transmit $RC_l$  packets in each time slot. With the increase in capacities, our problem can be rewritten as follows:

\textbf{Schedule($R$):}
\begin{align}
Max &\sum_{m,k:k\in V(m)}X_{mk} \label{RSchedule1} \\
s.t.&\sum_{k:k\in V(m)}X_{mk}\leq 1, \forall m\in \mathcal{M},\label{RSchedule2}\\
	&\sum_{m,k:(l,t)\in k} X_{mk} \leq RC_l, \forall l\in\mathcal{L},t\in\{1,2,\dots\},  \label{RSchedule3}\\
	& X_{mk}\geq 0, \forall  m\in \mathcal{M},k\in V(m). \label{RSchedule4}
\end{align}

To evaluate the performance of online policies, we define a competitive ratio that incorporates the increase in capacities:

\begin{definition}
\label{def:compratio}
Given a sequence of packet arrivals, let $\Gamma_{opt}$ be the optimal value of $\sum_{mk:k\in V(m)}X_{mk}$ in \textbf{Schedule}, and $\Gamma_{\eta}(R)$ be the number of packets that are delivered under an online policy $\eta$ when the link capacities are increased by $R$ times. The online policy $\eta$ is  said to be \emph{$(R,\rho)$-competitive} if $\Gamma_{opt}/\Gamma_{\eta}(R)\leq \rho$, for any sequence of packet arrivals.
\end{definition}

\section{An Online Algorithm and Its Competitive Ratio}
\label{section:algorithm}
\subsection{Algorithm Description}
In this section, we propose an online policy based on primal-dual method and analyze the competitive ratio. We first note that the dual problem of \textbf{Schedule} is:
\\
\\
\textbf{Dual}:
\begin{align}
	Min &\sum_{m}\alpha_m+\sum_{l,t} C_l \beta_{lt},  \label{dual1}\\
	s.t.\ & \alpha_m+\sum_{l,t:(l,t) \in k} \beta_{lt} \geq 1, \forall m\in\mathcal{M},k\in V(m) \label{dual2}\\
		& \alpha_m \geq 0, \forall m, \label{dual3}\\
		& \beta_{lt} \geq 0, \forall l,t, \label{dual4}
	\end{align}
where $\alpha_m$ is the Lagrange multiplier corresponding to constraint (\ref{Schedule2}), and $\beta_{lt}$ is the Lagrange multiplier corresponding to constraint (\ref{Schedule3}). 

By the Weak Duality Theorem, we have the following lemma:

\begin{lemma} 
\label{lemma:weak_duality}
Given any vectors of $\{\alpha_m\}$ and $\{\beta_{lt}\}$ that satisfy the constraints (\ref{dual2})--(\ref{dual4}), we have $\sum_{m}\alpha_m+\sum_{(l,t)} C_l \beta_{lt} \geq \Gamma_{opt}$.
\end{lemma}

We now introduce our online algorithm. Our algorithm constructs $\{X_{mk}\},\{\alpha_m\}, \{\beta_{lt}\}$ simultaneously while ensuring they satisfy all constraints in \textbf{Schedule($R$)} and \textbf{Dual}. Initially, it sets $\beta_{lt}\equiv 0$. When a packet $m$ arrives, the algorithm finds the valid schedule $k^*$ that has the largest $(1-\sum_{l,t:(l,t)\in k}\beta_{lt})$ among all $k\in V(m)$. If $1-\sum_{l,t:(l,t)\in k^*}\beta_{lt}\leq 0$, then the algorithm drops packet $m$ and sets $\alpha_m=0$ and $X_{mk}=0$, for all $k\in V(m)$. On the other hand, if $1-\sum_{l,t:(l,t)\in k^*}\beta_{lt}>0$, packet $m$ is transmitted using the valid schedule $k^*$. Our algorithm sets $X_{mk^*}=1$, $\alpha_m = 1-\sum_{l,t:(l,t)\in k^*}\beta_{lt}$, and updates $\beta_{lt}$ as $\beta_{lt} = \beta_{lt}(1+\frac{1}{C_{l}})+\frac{1}{(d_{l}-1)C_{l}}$ for all $(l,t) \in k^*$, where $d_l$ is chosen to be $(1+1/C_{l})^{RC_{l}}$. The complete policy is shown in Algorithm~\ref{Alg:PD}.

\begin{algorithm}
\caption{Online Algorithm with Variable $R$}
\label{Alg:PD}
\begin{algorithmic}[1]
  \STATE Initially, $\alpha_m\leftarrow 0$, $\beta_{lt}\leftarrow 0$, $X_{mk}\leftarrow 0$.
  \label{step:init}
  
  \STATE $d_l\leftarrow (1+1/C_{l})^{RC_{l}}, \forall l$.
  \label{step:d}
  
  \FOR{each arriving packet $m$}
 	\STATE $k^* \leftarrow \argmax_k {(1-\sum_{(l,t)\in k}\beta_{lt})}$ \label{step:k*}
  
  \IF{$(1-\sum_{(l,t)\in k^*}\beta_{lt})>0 $} \label{step:checkalpha}
 	\STATE $\alpha_m \leftarrow  {(1-\sum_{(l,t)\in k^*}\beta_{lt})}$ \label{step:alpha}
  
  	\STATE $\beta_{lt} \leftarrow \beta_{lt}(1+\cfrac{1}{C_{l}})+\cfrac{1}{(d_{l}-1)C_{l}}, \ (l,t)\in k^*$   	\label{step:beta}
  	
  	\STATE $X_{mk^*}\leftarrow 1.$   
  	\label{step:X}
  		
  	\STATE Transmit packet $m$ using valid schedule $k^*$.
  	\ELSE
  	\STATE Drop packet $m$.

  \ENDIF
  \ENDFOR
\end{algorithmic}
\end{algorithm}

\subsection{Complexity of the Algorithm}
In step \ref{step:k*}, the algorithm finds the valid schedule $k^*$ that maximizes $(1-\sum_{l,t:(l,t)\in k}\beta_{lt})$. We now show that this step can be completed in polynomial time by dynamic programming. 
Before presenting the algorithm, some new notations are given as follows. We say that packet $m$ joins the network at the beginning of time slot $a_m$, and specifies its deadline as $f_m$. Its source node and destination node are $s_m$ and $d_m$, respectively. Therefore, a valid schedule for $m$ is one that can deliver a packet from node $s_m$ to node $d_m$ between time slots $a_m$ and $f_m$.

Let $\Theta(n,\tau)$ be the smallest value of $\sum_{l,t:(l,t)\in k}\beta_{lt}$ among all schedules that can deliver a packet from node $s_m$ to node $n$ between time slots $a_m$ and $\tau$. $\Theta(n,\tau)=\infty$ if there is no schedule that delivers a pacekt from $s_m$ to $d_m$ between time slots $a_m$ and $\tau_m$. Step \ref{step:k*} of Alg. \ref{Alg:PD} is then equivalent to finding the valid schedule that achieves $\sum_{l,t:(l,t)\in k}\beta_{lt}=1-\Theta(d_m, f_m)$. Since packet $m$ arrives at the beginning of time slot $a_m$, or, equivalently, at the end of time slot $a_m-1$, we set $\Theta(s_m, a_m-1)=0$ and $\Theta(n, a_{m}-1)=\infty$ for $\forall n \neq s_m$. 

There are only two different ways to deliver a packet to node $n$ by the end of time slot $\tau$: The first is to deliver the packet to $n$ by time slot $\tau-1$, in which case $\sum_{l,t:(l,t)\in k}\beta_{lt}=\Theta(n,\tau-1)$. The second is to deliver the packet to one of $n$'s neighbors, say, node $q$, by time slot $\tau-1$, and then forward the packet along the link $l_{qn}$ from $q$ to $n$ at time slot $\tau$. In this case, $\sum_{l,t:(l,t)\in k}\beta_{lt}=\Theta(q,\tau-1)+\beta_{l_{qn}\tau}$. Therefore, we have
\[
\Theta(n,\tau)=min\left \{ \begin{array}{l}
\Theta(n,\tau-1),\\
\Theta(q, \tau-1)+\beta_{l_{qn}\tau}, \mbox{$q$ is a neighbor of $n$.}\end{array}\right.
\]

Based on the above recursive equation, we design an algorithm for computing $\Theta(n,\tau)$. The detailed algorithm is shown in Algorithm \ref{Alg:DP}, where we also use $Sch(n,\tau)$ to denote the schedule that achieves $\Theta(n,\tau)$.

\begin{algorithm}
\caption{Dynamic Programming}
\label{Alg:DP}
\begin{algorithmic}[1]
  \FOR {each arriving packet $m$}
   \STATE $\Theta(s_m, a_m-1)\leftarrow 0$
  \STATE $\Theta(n, a_m-1)\leftarrow \infty,\forall n\neq s_m$
  \STATE $Sch(n,a_m-1)\leftarrow\phi, \forall n$
  	\FOR {$\tau=a_m$ \TO $f_m$}
  		\FOR {node $n$}
  			\STATE $\Theta(n,\tau)\leftarrow \Theta(n,\tau-1)$
  			\STATE $Sch(n,\tau)\leftarrow Sch(n, \tau-1)$
  			\FOR {node $n$'s neighbor $q$}
  			\IF {$\Theta(q,\tau-1)+\beta_{l_{qn}\tau}<\Theta(n,\tau)$}
  			\STATE $\Theta(n,\tau)\leftarrow\Theta(q,\tau-1)+\beta_{l_{qn}\tau}\}$
  			\STATE $Sch(n,\tau)\leftarrow Sch(q, \tau-1)\cup \{(l_{qn}, \tau)\}$
  			\ENDIF
  			\ENDFOR
  		\ENDFOR
  	\ENDFOR
  \ENDFOR
\end{algorithmic}
\end{algorithm}

%We aim to find  $\min \sum_{(l,t)\in k}\beta_{lt}$ among all schedules $k \in V(m)$. Consider a $\beta_{lt}$ with $(l,t) \in k$ here. At each link $l$, the packet is either transmitted on link $l$ at time $t$ or being transmitted on link $l$ before time $t$ and stay un-transmitted during time $t$. 
%Assume the last transmitted link is $l$ up to time $t$, let $\Theta(l,t)$ be the value of $\min \sum_{(l,t)\in k}\beta_{lt}$. Let $s_l$ be the source node of link $l$. Let $(l^p, t-1) \in k$ and $l^p$ connects to node $s_l$. Then $\Theta(l,t)=\min \{ \Theta(l,t-1), \Theta(l^p,t-1)+\beta_{lt}\}$. We repeat this step for each link on the route for this packet. The algorithm is shown in Algorithm \ref{Alg:DP}. 

%\begin{algorithm}
%\caption{Dynamic Programming}
%\label{Alg:DP}
%\begin{algorithmic}[1]
%  \STATE $\beta_{lt}$ are calculated by Algorithm \ref{Alg:PD}
%  
%  \FOR {each arriving packet $m$}
%  	\FOR {$l:(l,t)\in k$} 
%  		\FOR {$t:(l,t)\in k$}
%  			\STATE $\Theta(l,t)=\min \{ \Theta(l,t-1), \Theta(l^p,t-1)+\beta_{lt}\}$.
%  			\IF {$\Theta(l,t)= \Theta(l,t-1)$}
%  				\STATE $keep(l,t)=0$
%  			\ELSE
%  				\STATE $keep(l,t)=1$
%  			\ENDIF
%  		\ENDFOR
%  	\ENDFOR
%  \FOR {$l$ from destination node to origin}
%  	\IF {$keep(l,t)=1$}
%  		\STATE output $l$, $t$
%  	\ENDIF
%  \ENDFOR
%  \ENDFOR
%
%\end{algorithmic}
%\end{algorithm}
In Alg. \ref{Alg:DP}, the inequality $\Theta(q,\tau-1)+\beta_{l_{qn}\tau}<\Theta(n,\tau)$ is only evaluated once for any link and time slot. Let $E$ be the number of links in the system. Suppose the number of links is larger than the number of nodes, and $f_m-a_m+1\leq T$, for all $m$, then the complexity of Alg. \ref{Alg:DP} is $O(ET)$.

\subsection{Competitive Ratio Analysis}
Before analyzing the performance of Algorithm \ref{Alg:PD}, we first establish a basic property of the values of $\beta_{lt}$.
\begin{lemma}
\label{lemma:d}
Let ${\beta_{lt}}[n]$ be the value of $\beta_{lt}$ after $n$ packets are scheduled to use link $l$ at time $t$. Then,
\begin{align}
\beta_{lt}[n] = (\frac{1}{d_l-1})(d_l^{n/{RC_j}}-1). \label{beta}
\end{align} 
\end{lemma}

\begin{IEEEproof}
First, note that the value of $\beta_{lt}$ is only changed when Algorithm~\ref{Alg:PD} uses link $l$ at time $t$ to transmit a packet. Therefore, the value of $\beta_{lt}$ only depends on the number of packets that are scheduled to use link $l$ at time $t$.

We then prove \eqref{beta} by induction. Initially, when $n=0$, $\beta_{lt}[0]=0=(\frac{1}{d_l-1}) (d_l^0-1)$
and \eqref{beta} holds.

Suppose \eqref{beta} holds for the first $n$ packets. When the $(n+1)$-th packet is scheduled for link $l$ at time $t$, we have

\begin{align*}
	{\beta_{lt}}[n+1]
   =&{\beta_{lt}}[n](1+\frac{1}{C_l})+\frac{1}{(d_l-1)C_l} \\	
   =& \cfrac{1}{(d_l-1)}(d_l^{n/RC_l}-1)(1+\frac{1}{C_l})+\cfrac{1}{(d_l-1)C_l} \\	
   =&\frac{1}{d_l-1}[d_l^{n/RC_l}(1+\frac{1}{C_l})-1]
\end{align*}

We select $d_l=(1+\frac{1}{C_l})^{RC_l}$, and therefore 
\begin{align*}
	{\beta_{lt}}[n+1]
	= \frac{1}{(d_l-1)} [ d_l^{(n+1)/RC_l} -1 ],
\end{align*}	
and \eqref{beta} still holds for $n+1$. Thus, by induction, \eqref{beta} holds for all $n$.
\end{IEEEproof}

We now establish the competitive ratio of Algorithm~\ref{Alg:PD}.
\begin{theorem}
\label{theorem:compratio}
Let $C_{min}:=\min C_l$, $d_{min}:=(1+1/C_{min})^{RC_{min}}$, and $L$ be the longest path between a source node and a destination node, that is, all valid schedules have $|k|\leq L$, for all $m\in \mathcal{M}, k\in V(m)$. Algorithm \ref{Alg:PD} produces solutions that satisfy all constraints in \textbf{Schedule($R$)} and \textbf{Dual}. Moreover,  Algorithm \ref{Alg:PD} is $(R,1+\frac{L}{d_{min}-1} )$-competitive, which converges to $(R,1+\frac{L}{e^R-1} )$-competitive, as $C_{min}\rightarrow\infty$. 
\end{theorem}

\begin{IEEEproof}
First, we show that the dual solutions $\{\alpha_m \}$ and $\{ \beta_{lt} \}$ satisfy constraints \eqref{dual2} to \eqref{dual4}. 
Initially, we have $\beta_{lt}=0$. By Lemma \ref{lemma:d}, $\beta_{lt} \geq 0$ holds. Since step \ref{step:alpha} is only used when $(1-\sum_{(l,t)\in k^*}\beta_{lt})>0$, $\alpha_m\geq0$ holds. From step \ref{step:k*} and \ref{step:alpha}, we know that $\alpha_m+\sum_{(l,t) \in k} \beta_{lt} \geq (1-\sum_{(l,t) \in k} \beta_{lt}) +\sum_{(l,t) \in k} \beta_{lt}=1$. Thus \eqref{dual2} to \eqref{dual4} hold.

Next, we show $\{X_{mk}\}$ satisfies constraints \eqref{RSchedule2} to \eqref{RSchedule4}. By step \ref{step:k*}, the algorithm picks at most one schedule $k^*$ for packet $m$, constraint \eqref{RSchedule2} holds. 
With Lemma \ref{lemma:d}, $\beta_{lt}=1$ when $RC_l$ packets use link $l$ at time $t$. Since a valid schedule including $(l,t)$ will be chosen for packet $m$ only when $(1-\sum_{(l,t)\in k^*}\beta_{lt})>0$, all $(l,t)$ in the chosen valid schedule must have $\beta_{lt}<1$, and therefore the number of packets transmitted over link $l$ at time $t$ must be less than $RC_l$. Thus, at any time $t$, there are at most $RC_l$ packets using link $l$. Constraint \eqref{RSchedule3} holds. 
By initialization and step \eqref{step:X}, constraint \eqref{RSchedule4} holds.

We derive the ratio between $\sum_{m}\alpha_m+\sum_{(l,t)} C_l \beta_{lt}$ and $\sum_{mk}X_{mk}$. Initially, both are equal to 0. We consider the increasing amount for both when a new packet $m$ arrives at the network. We use $\Delta P(R)$ to denote the change of $\sum_{mk}X_{mk}$, and $\Delta D$ to denote the change of $\sum_{m}\alpha_m+\sum_{(l,t)} C_l \beta_{lt}$.

If packet $m$ is dropped, both $\Delta P(R)$ and $\Delta D$ are 0. 
If packet $m$ is accepted and transmitted using valid schedule $k^*$, we have $X_{mk^*}=1$. Thus, $\Delta P(R)=1$. On the other hand, $\Delta D$ is increased as:

\begin{align*}
\Delta D 
=& \alpha_m+\sum_{(l,t)\in k^*} C_l \Delta \beta_{lt} \\
=& (1-\sum_{(l,t)\in k^*} \beta_{lt}) +\sum_{(l,t)\in k^*} (\beta_{lt}+\frac{1}{(d_l-1)C_l} )\\
=& 1+ \sum_{(l,t) \in k^*} {\frac{1}{(d_l-1)}}\leq  1+\frac{L}{d_{min}-1} \\
\end{align*}

Therefore, for each packet arrival, the ratio between $\Delta D$ and $\Delta P(R)$is no larger than $1+\frac{L}{d_{min}-1}$ if $\Delta D>0$. When the algorithm terminates, we have $\frac{\sum_{m}\alpha_m+\sum_{(l,t)} C_l \beta_{lt}}{\sum_{mk}X_{mk}}\leq 1+\frac{L}{d_{min}-1}$.
By Lemma \ref{lemma:weak_duality}, $\frac{\Gamma_{opt}}{\sum_{mk}X_{mk}}\leq 1+\frac{L}{d_{min}-1}$, and the competitive ratio of Algorithm \ref{Alg:PD} is $(R, 1+\frac{L}{d_{min}-1})$. When $C_{min}\rightarrow\infty$, $d_{min}=(1+\frac{1}{C_{min}})^{RC_{min}}\rightarrow e^R$, and the competitive ratio of Algorithm~\ref{Alg:PD} converges to $(R, 1+\frac{L}{e^R-1})$.
\end{IEEEproof}

There are several important implications of Theorem~\ref{theorem:compratio}. 
First, without increasing capacity, that is, when $R=1$, the competitive ratio of our policy is $(1, O(L))$. In comparison, the online algorithm proposed in the recent work \cite{multihoponline} focuses on the special case of $R=1$ and has a competitive ratio of $(1, O(L\log L))$. Therefore, our algorithm is asymptotically better than the online algorithm in \cite{multihoponline}. 
Second, this theorem allows us to quantify the amount of capacity needed to a certain performance guarantee. Suppose the optimal solution to \textbf{Schedule} indeed delivers all packets. In order to guarantee that $1-\frac{1}{\theta}$ of the packets are transmitted to their destinations before their deadlines, Theorem~\ref{theorem:compratio} states that we only need to increase all link capacities by $R_{\theta}$ times so that $1+\frac{L}{e^{R_{\theta}}-1}\leq 1/(1-\frac{1}{\theta})=1+\frac{1}{\theta-1}$. Therefore, we have $R_{\theta}= \ln {(L(\theta-1)+1)}\leq \ln L+\ln \theta$. For example, if we are required to deliver $99 \% $ of the packets and the longest path consists of 10 hops, then we need to increase capacity by 6.9 times.

\section{A Theoretical Lower Bound for Competitive Ratio}	\label{section:lower bound} 
In Section \ref{section:algorithm}, we showed that our policy is $(R, 1+\frac{L}{e^R-1})$-competitive. In this section, we will establish a lower bound for the competitive ratio of online policies.

\begin{theorem}
\label{theorem:bound}
Any online algorithm cannot be better than $(R, 1+\frac{L-2e^R}{(L+1)e^R-L})$-competitive.
\end{theorem}
\begin{IEEEproof}
We design a network as shown in Fig \ref{fig:bound1}. We start to construct the network from an up-link tree, which is shown as the white nodes in Fig \ref{fig:bound1}. Root is marked as node $D$ and it is the destination of all packets. There are $N$ levels of non-root nodes with $N$ nodes in each level. Each node is connected to one node in the next level. Nodes do not share parent except the $N$-th level nodes share the same root node. 
At the $j$-th level, where $1 \leq j \leq N$, there are $N \choose N+1-j$ extra nodes, which is shown as the black nodes in Fig \ref{fig:bound1}, with each node connecting to an unique set of $N+1-j$ nodes in this level. For example, there is one black node connected to all white nodes in level 1, and there are $N$ black nodes connected to white nodes in level 2, where each of these black nodes is connected all but one white nodes in level 2. Likewise, there are $N\choose N-2$ black nodes connected to white nodes in level 3, with each black node connected to $N-2$ white nodes in level 3, and no two black nodes are connected to the same subset of white nodes.

\begin{figure}
	\begin{center}
	\includegraphics[width=3.5in]{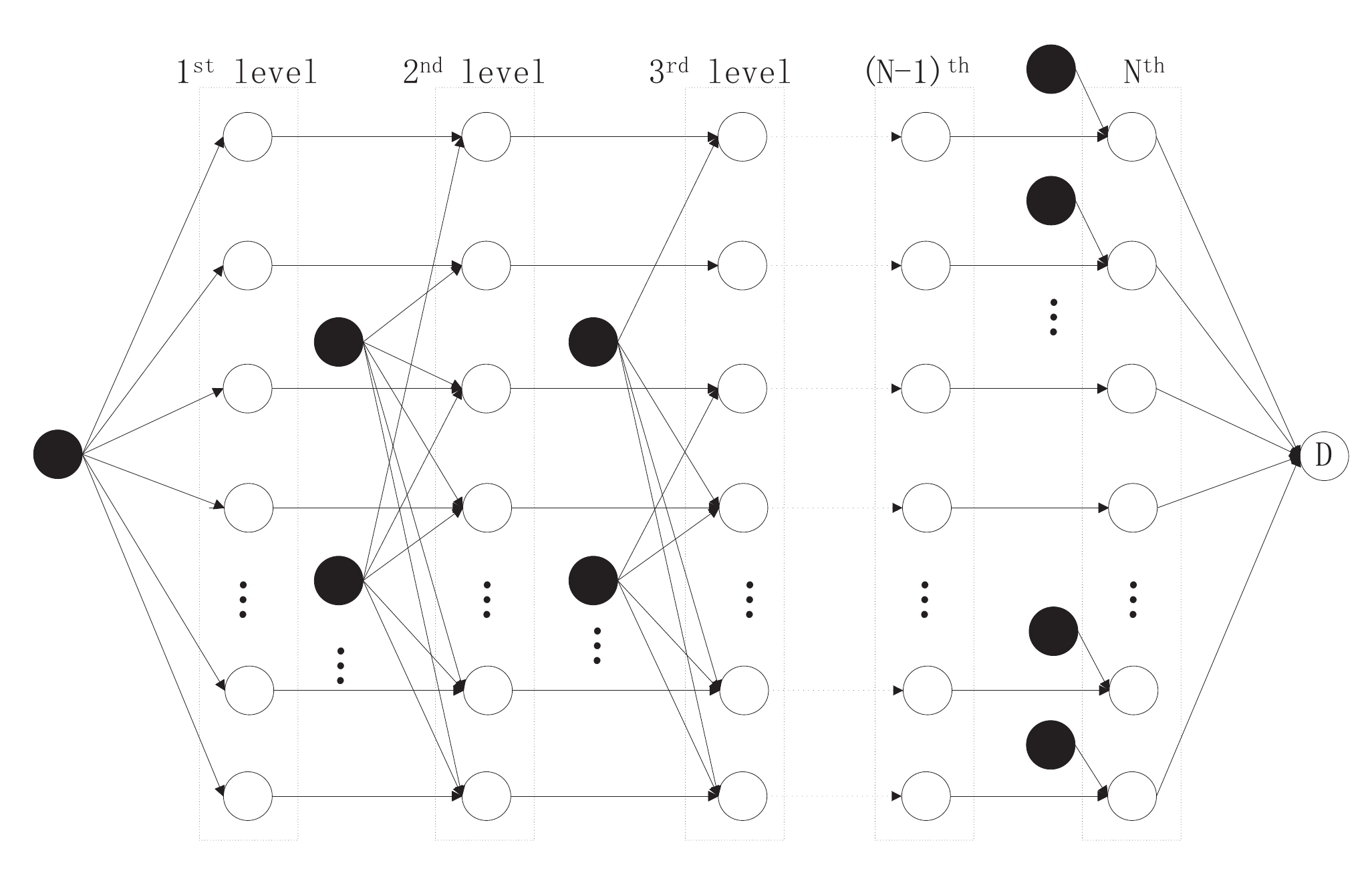}
	\caption{Network topology for lower bound analysis}
	\label{fig:bound1}
	\end{center}
\end{figure}

Next, we describe packet arrivals. Packets only arrive at black nodes. Of all black nodes connected to the same level of white nodes, only one black node has packet arrival. Let $\mathcal{W}_j$ be the set of white nodes in $j$-th level which connects to the black node with packet arrivals. The black nodes with packet arrivals are chosen such that all nodes in $\mathcal{W}_{j+1}$ are connected to those in $\mathcal{W}_j$. 
Fig \ref{fig:bound2} is a simplified network of Fig \ref{fig:bound1}, where we omit the black nodes with no packet arrival and marked each black node with a number from 1 to $N$.

\begin{figure}
	\begin{center}
	\includegraphics[width=3.5in]{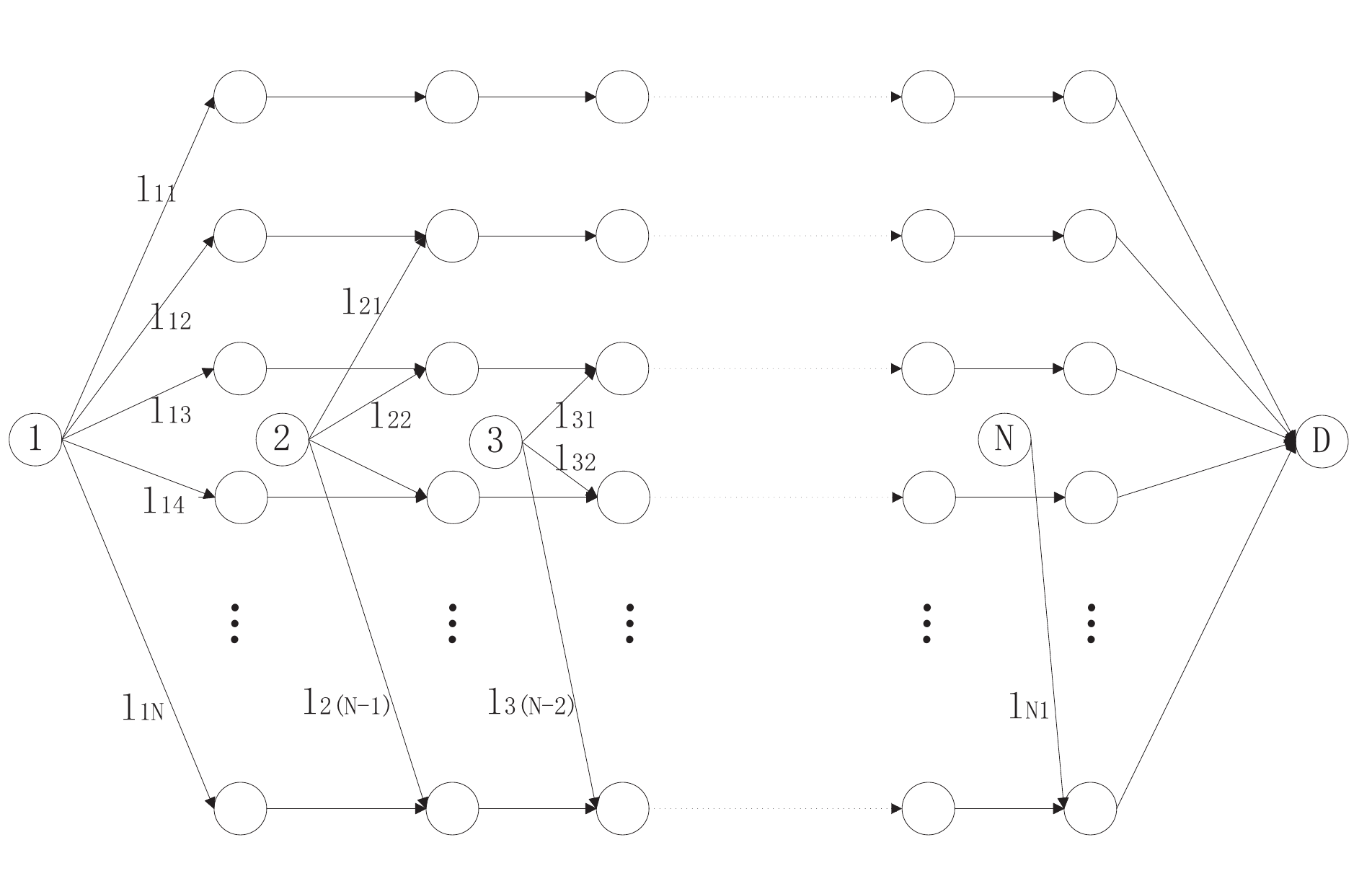}
	\caption{Simplified network topology for lower bound analysis}
	\label{fig:bound2}
	\end{center}
\end{figure}

Packets arrive at nodes $1,2,...,N$. Their destination is node $D$. Each link in the network has capacity $C$. At the beginning of time slot $1$, there are $C$ packets arriving at node $1$. Node $1$ is connected to $N$ links: $l_{11}$, $l_{12}$, $\cdots$, $l_{1N}$. At the beginning of time slot $2$, there are $C$ packets arriving at node $2$. Node $2$ is connected to $N-1$ links: $l_{21}$, $l_{22}$, $\cdots$, $l_{2(N-1)}$. Similarly for nodes $3$, $4$, $\cdots$. At the beginning of time $N$, there are $C$ packets arriving at node $N$. The deadline of all packets is $N+1$. Node $N$ is connected only to link $l_{N1}$.

When one knows which black nodes have packet arrivals, the offline optimal algorithm is to transmit the first $C$ packets through link $l_{11}$ and the following links, the second $C$ packets through link $l_{21}$ and the following links, $\dots$, and the $N$-th $C$ packets through link $l_{N1}$ and the following link. The total number of delivered packets is $NC$.

Next we consider the online algorithm when all links' capacity is increased by $R$ times. Since online policies do not know which black nodes will have packet arrivals, the optimal online policy is to distribute packets evenly among all connected links. That is, at time $1$, each of links $l_{1i}$, $i=1,2, \cdots, N$, transmit $C/N$ packets. At time $2$, each of link $l_{2i}$, $i=1,2, \cdots, (N-1)$, transmits $C/(N-1)$ packets. At time $K$, link $l_{Ki}$, $i=1,2, \cdots, (N-K+1)$, transmits $C/(N-K+1)$ packets. For simplicity, we call the routes from node $1$ to node $D$ through $l_{1i}$ route $r_i$. If all packets arrive at node $K$ are accepted, routes $r_{i}$, $i=K, K+1, \cdots, N$ have the same load on each link.
When any link on a single route reaches its capacity, the route cannot be used for future arrival packets. Suppose the route gets over-loaded at time $K+1$, that is, packets arrive at node $K$ are accepted and packets arrive at node $K+1$ are not fully accepted.  The maximum load of a single link on route $r_N$ is at most $\frac{C}{N}+\frac{C}{N-1}+\cdots+\frac{C}{N-K+1}$ and  at least $\frac{C}{N}+\frac{C}{N-1}+\cdots+\frac{C}{N-K}$.
We then have:
$$C(\frac{1}{N}+\frac{1}{N-1}+\frac{1}{N-2}+\ldots+\frac{1}{N-K+1}) \leq RC,$$ and $$C(\frac{1}{N}+\frac{1}{N-1}+\frac{1}{N-2}+\ldots+\frac{1}{N-K}) \geq RC.$$

Since
$$\int_{N-K+1}^{N+1} \frac{1}{x} dx < (\frac{1}{N}+\frac{1}{N-1}+\frac{1}{N-2}+\ldots+\frac{1}{N-K+1}),$$
and
$$\int_{N-K-1}^{N} \frac{1}{x} dx  >(\frac{1}{N}+\frac{1}{N-1}+\frac{1}{N-2}+\ldots+\frac{1}{N-K}).$$

We have: 
$$\log (N+1) - \log (N-K+1) = \log {\frac{N+1}{N-K+1}}< R, $$
and
$$\log (N) - \log (N-K-1) = \log {\frac{N}{N-K-1}} > R. $$

Then we can derive the value of $K$ as:
$N- \frac{N}{e^R}-1 \leq K \leq N+1 - \frac{N+1}{e^R}$. 
The total number of accepted packets is in the range $((N- \frac{N}{e^R}-1)C, (N+2 - \frac{N+1}{e^R})C)$. 

Thus the competitive ratio of an online policy is at best $(R,\frac{N}{N+2- \frac{N+1}{e^R}})$. In Fig. \ref{fig:bound1}, the longest path in the network is between the leftmost black node and the sink, which has length $L=N+1$. The competitive ratio can then be rewritten as $(R, 1+\frac{L-2e^R}{(L+1)e^R-L})$.
\end{IEEEproof}

Let us once again consider the scenario where online policies need to guarantee to deliver at least $1-\frac{1}{\theta}$ as many packets as the optimal solution. Theorem~\ref{theorem:bound} states that any online policy needs to increase its link capacities by at least $R_{\theta}$ times so that $1+\frac{L-2e^{R_\theta}}{(L+1)e^{R_\theta}-L}\leq 1+\frac{1}{\theta-1}$. Solving this equation, and we have $R_{\theta}$ needs to be at least $\ln L+\ln \theta-\ln(L+2\theta-1)$. In comparison, our policy only needs to increase link capacities by $(\ln L+\ln \theta)$ times to ensure the delivery of $1-\frac{1}{\theta}$ as many packets as the optimal solution. Therefore, the capacity requirement of our policy is at most $\ln(L+2\theta-1)$ away from the lower bound. Suppose we fix the ratio between $L$ and $\theta$, and let them both go to infinity, then we have $(\ln L+\ln \theta)/(\ln L+\ln \theta-\ln(L+2\theta-1))\rightarrow 2$. Therefore, when both $L$ and $\theta$ are large, our policy at most requires twice as much capacity as the theoretical lower bound.

\section{An Order-Optimal Online Policy with Fixed $R=1$}
\label{section:modified alg}

We have shown that Alg. \ref{Alg:PD} is $(R,1+\frac{L}{d_{min}-1} )$-competitive. Without increasing link capacity, i.e, $R=1$, the algorithm is $(1, 1+\frac{L}{e-1})$-competitive, as $C_{min}\rightarrow\infty$. While the competitive ratio of Alg. \ref{Alg:PD} is an order better than that of the online policy in the previous work \cite{multihoponline}, it still fails to achieve the theoretical bound of $(1, O(\log L))$-competitive. In this section, we propose another online algorithm and prove that it achieves the theoretical bound when $R=1$.

\subsection{Algorithm Description}
Similar to the design of Alg. \ref{Alg:PD}, we aim to design an algorithm that constructs $\{X_{mk}\},\{\alpha_m\}, \{\beta_{lt}\}$ while ensuring they satisfy all constraints in \textbf{Schedule} and \textbf{Dual}. The algorithm is described in Alg. \ref{Alg:ModifiedPD}. One can see that Alg. \ref{Alg:ModifiedPD} is very similar to Alg. \ref{Alg:PD}, and their only difference lie in the update rules for $\beta_{lt}$. Specifically, let $\beta_{lt}[n]$ be the value of  $\beta_{lt}$ when link $l$ serves a total number of $n$ packets at time $t$. Then Alg. \ref{Alg:ModifiedPD} chooses the value of $\beta_{lt}[n]$ as:
\begin{align}
\beta_{lt}[n]=\left\{
    \begin{array}{ll}
		\cfrac{1}{L(e^{\frac{1}{\ln L +1}}-1)}(e^{\frac{n}{C_l}}-1), \mbox{if $n \leq \frac{C_l}{\ln L+1}$}; \\
		e^{(\frac{n}{C_l}-1)(\ln L +1)}, \mbox{if $n \geq \frac{C_l}{\ln L+1}$.}
    \end{array} \label{beta:Mofified}
\right.
\end{align}

\begin{figure}
	\begin{center}
	\includegraphics[width=3.5in]{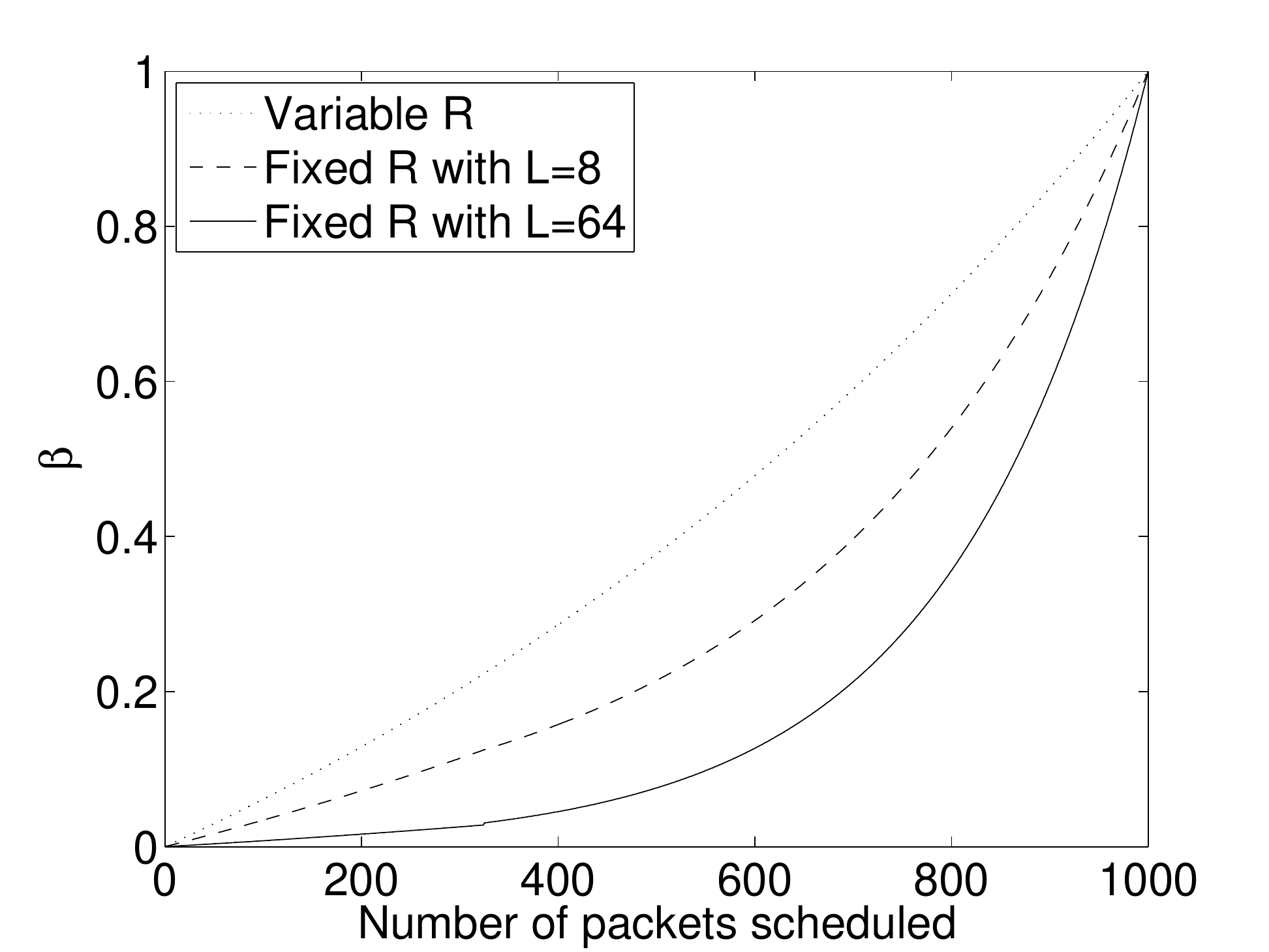}
	\caption{Values of $\beta_{lt}$ under different policies.}
	\label{fig:compareR}
	\end{center}
\end{figure}

To illustrate the difference in $\beta_{lt}$, we plot the values of $\beta_{lt}[n]$ for a link with $C_l=1000$ under the two policies in Fig. \ref{fig:compareR}, where we consider the two cases $L=8$ and $L=64$ for Alg. \ref{Alg:ModifiedPD}. As can be shown in the figure, when $n$ is small, Alg. \ref{Alg:ModifiedPD} increases the value of $\beta_{lt}$ much slower than Alg. \ref{Alg:PD} does. Moreover, Alg. \ref{Alg:ModifiedPD} increases $\beta_{lt}$ slower when $L$ is larger. Recall that both Alg. \ref{Alg:PD} and Alg. \ref{Alg:ModifiedPD} only schedule a packet when $\max_k {(1-\sum_{(l,t)\in k}\beta_{lt})}>0$, or, equivalently, $\min_k {\sum_{(l,t)\in k}\beta_{lt}}<1$. By increasing $\beta_{lt}$ slower when $n$ is small, Alg. \ref{Alg:ModifiedPD} ensures that more packets with long routes can be accepted, especially when the network is lightly loaded.

\begin{algorithm}
\caption{Online Algorithm with Fixed $R=1$}
\label{Alg:ModifiedPD}
\begin{algorithmic}[1]
  \STATE Initially, $\alpha_m\leftarrow 0$, $\beta_{lt}\leftarrow 0$, $X_{mk}\leftarrow 0$.
  \label{stepM:init}

  \FOR{each arriving packet $m$}
 	\STATE $k^* \leftarrow \argmax_k {(1-\sum_{(l,t)\in k}\beta_{lt})}$ \label{stepM:k*}
  
  \IF{$(1-\sum_{(l,t)\in k^*}\beta_{lt})>0 $} \label{stepM:checkalpha}
 	\STATE $\alpha_m \leftarrow  {(1-\sum_{(l,t)\in k^*}\beta_{lt})}$ \label{stepM:alpha}
 	\FOR{each $(l,t)\in k^*$}
 	\IF{total number of packets $n$ at time $t$ on link $l$: $n\leq \frac{C_l}{\ln L+1}$}
  		\STATE $\beta_{lt} \leftarrow \cfrac{1}{L(e^{\frac{1}{\ln L +1}}-1)}(e^{\frac{n}{C_l}}-1),$ 	\label{stepM:beta1}
  		\ELSE
  		\STATE $\beta_{lt} \leftarrow e^{(\frac{n}{C_l}-1)(\ln L +1)}$ 	\label{stepM:beta2}
  	\ENDIF
  	\ENDFOR
  	\STATE $X_{mk^*}\leftarrow 1.$   
  	\label{stepM:X}
  		
  	\STATE Transmit packet $m$ using valid schedule $k^*$.
  	\ELSE
  	\STATE Drop packet $m$.

  \ENDIF
  \ENDFOR
\end{algorithmic}
\end{algorithm}

\subsection{Competitive Ratio Analysis}

We now prove that Alg. \ref{Alg:ModifiedPD} achieves the theoretical bound in \cite{multihoponline} by being $(1, O(\log L))$-competitive.

\begin{lemma}
\label{lemma:modified}
Let $C_{min}:=\min C_l$. In Algorithm~\ref{Alg:ModifiedPD}, each time a new packet is scheduled, the ratio between the change of \textbf{Schedule} and \textbf{Dual} is bounded by $2(\ln L +1)+ \frac{B}{C_{min}}$, where the value of $B$ is independent of $C_{min}$.
\end{lemma}

\begin{IEEEproof}
If a new packet is admitted to the network, the increasing amount of \textbf{Dual} is

\begin{align*}
\Delta D 
=& \alpha_m+\sum_{(l,t)\in k^*} C_l \Delta \beta_{lt} \\
=& 1+\sum_{(l,t)\in k^*} (C_l \Delta \beta_{lt} - \beta_{lt}) \\
\end{align*} 

We define $\beta(x)$ as
\begin{align}
\beta(x)=\left\{
    \begin{array}{ll}
		\cfrac{1}{L(e^{\frac{1}{\ln L +1}}-1)}(e^{x}-1), \mbox{if $x \leq \frac{1}{\ln L+1}$}; \\
		e^{(x-1)(\ln L +1)}, \mbox{if $x \geq \frac{1}{\ln L+1}$.}
    \end{array} 
\right.
\end{align}
Note that $\beta_{lt}[n] = \beta(\frac{n}{C_l})$. By using Taylor Sequence, we then have 
\begin{align*}
\Delta{\beta_{lt}}[n]&:=\beta_{lt}[n+1]-\beta_{lt}[n]=\beta(\frac{n+1}{C_l})-\beta(\frac{n}{C_l})\\
&\leq \frac{1}{C_l}\beta'(\frac{n}{C_l})+\epsilon \frac{1}{C_l^2}\beta''(\frac{n}{C_l}),
\end{align*} 
for some bounded constant $\epsilon<\infty$, where $\beta'$ and $\beta''$ are the first and second derivative of $\beta$, respectively. We note that the function $\beta(x)$ is continuous for all $x$, and infinitely differentiable for all $x$ except at the point $x_0:=\frac{1}{\ln L+1}$. At the point $x_0$, we define $\beta'(x_0)=\lim_{x \rightarrow x_0^+}\beta'(x)$ and $\epsilon \beta''(x_0)=\lim_{x \rightarrow x_0^+}\epsilon  \beta''(x)$. This ensures that the above inequality still holds.

By \eqref{beta:Mofified} we know that $n \leq\frac{C_l}{\ln L+1}$ if and only if $\beta_{lt}[n]\leq\frac{1}{L}$.

If $x=\frac{n}{C_l} \leq\frac{1}{\ln L+1}$,  then $\beta'(x)=\beta''(x) = \frac{e^x}{L(e^{\frac{1}{\ln L +1}}-1)}$. We have:

\begin{align*}
& C_l \Delta \beta_{lt}[n] - \beta_{lt}[n]   \\
\leq &\frac{C_l ( \frac{1}{C_l} e^{\frac{n}{C_l}})  + \epsilon (\frac{1}{C_l})^2 e^{\frac{n}{C_l}})-(e^{\frac{n}{C_l}}-1)}
{L(e^{\frac{1}{\ln L +1}}-1)} \\
\leq & \frac{1+\epsilon \frac{1}{C_l} e^{\frac{n}{C_l}}}{L(1+\frac{1}{\ln L +1}-1)} \\
\leq & \frac{\ln L+1}{L} (1+\epsilon \frac{1}{C_l} e) \\
\end{align*}

Let $B_1=\epsilon e\frac{\ln L+1}{L} $, then

\begin{align}
C_l \Delta \beta_{lt}[n] - \beta_{lt}[n]  
\leq \frac{\ln L+1}{L} + B_1 \frac{1}{C_{min}}, \label{deltaD1}
\end{align}
when $\frac{n}{C_l} \leq\frac{1}{\ln L+1}$.

On the other hand, If $x=\frac{n}{C_l} \geq\frac{1}{\ln L+1}$,  then $\beta'(x)=(\ln L +1)\beta(x)$ and $\beta''(x)=(\ln L +1)^2\beta(x)$. We have:
\begin{align*}
& C_l \Delta \beta_{lt}[n]  - \beta_{lt}[n]  \\
\leq & C_l [ \frac{\ln L+1}{C_l} \beta_{lt}[n]  + \epsilon (\frac{\ln L+1}{C_l})^2 \beta_{lt}[n]]-\beta_{lt}[n] \\
\leq & \ln L \cdot \beta_{lt}[n] +\frac{1}{C_l} \epsilon (\ln L+1)^2 \beta_{lt}[n] 
\end{align*}

Let $B_2=\epsilon (\ln L+1)^2 $, then

\begin{align}
& C_l \Delta \beta_{lt}[n]  - \beta_{lt}[n]  
\leq (\ln L+ B_2 \frac{1}{C_{min}}) \beta_{lt}[n], \label{deltaD2}
\end{align} 
when $\frac{n}{C_l} \geq\frac{1}{\ln L+1}$.

If packet $m$ is transmitted using valid schedule $k^*$, we have $X_{mk^*}=1$. Thus, $\Delta P=1$. On the other hand, $\Delta D$ is increased as:
\begin{align*}
\Delta D 
=	& 1+\sum_{(l,t): (l,t) \in k^*} {C_l \Delta \beta_{lt}  - \beta_{lt}} \\
\leq & 1+\sum_{(l,t): (l,t) \in k^*, \beta_{lt} \leq \frac{1}{L}} {C_l \Delta \beta_{lt}  - \beta_{lt}} \\ 
&\ \ +\sum_{(l,t): (l,t) \in k^*, \beta_{lt} \geq \frac{1}{L}} {C_l \Delta \beta_{lt}  - \beta_{lt}} 
\end{align*}

From \eqref{deltaD1} and \eqref{deltaD2} we have:
\begin{align*}
\Delta D 
\leq & 1+ \sum_{(l,t): (l,t) \in k^*, \beta_{lt} \leq \frac{1}{L}} (\frac{\ln L+1}{L} + B_1 \frac{1}{C_{min}} ) \\
&\ \ + \sum_{(l,t): (l,t) \in k^*, \beta_{lt} \geq \frac{1}{L}} ((\ln L+ B_2 \frac{1}{C_{min}}) \beta_{lt})
\end{align*}

From Algorithm \ref{Alg:ModifiedPD} step \ref{stepM:checkalpha} we know that $\sum \beta_{lt} \leq 1$, thus we have
\begin{align*}
\Delta D 
\leq & 1+(\ln L +1 + B_1 \frac{L}{C_{min}}) + (\ln L+B_2\frac{1}{C_{min}}) \\
=	 & 2 + 2\ln L +\frac{B_1+B_2}{C_{min}},
\end{align*}
and the proof is complete.
\end{IEEEproof}

\begin{theorem}
\label{theorem:Modifiedcompratio}
Algorithm \ref{Alg:ModifiedPD} produces solutions that satisfy all constraints in \textbf{Schedule} and \textbf{Dual}. Moreover, it is $(1,2(1+\ln L) )$-competitive, as $C_{min}\rightarrow\infty$.
\end{theorem}

\begin{IEEEproof}
First, we show that the dual solutions $\{\alpha_m \}$ and $\{ \beta_{lt} \}$ satisfy constraints \eqref{dual2} to \eqref{dual4}. 
Initially, we have $\beta_{lt}=0$. By \eqref{beta:Mofified}, $\beta_{lt} \geq 0$ holds. Since step \ref{stepM:alpha} is only used when $(1-\sum_{(l,t)\in k^*}\beta_{lt})>0$, $\alpha_m\geq0$ holds. From step \ref{stepM:k*} and \ref{stepM:alpha}, we know that $\alpha_m+\sum_{(l,t) \in k} \beta_{lt} \geq (1-\sum_{(l,t) \in k} \beta_{lt}) +\sum_{(l,t) \in k} \beta_{lt}=1$. Thus \eqref{dual2} to \eqref{dual4} hold.

Next, we show $\{X_{mk}\}$ satisfies constraints \eqref{Schedule2} to \eqref{Schedule4}. By step \ref{stepM:k*}, the algorithm picks at most one schedule $k^*$ for packet $m$, constraint \eqref{Schedule2} holds.
With \eqref{beta:Mofified}, when the number of packets on link $l$ at $t$ is $C_l$, we have $\beta_{lt}=1$. Also, since a packet is scheduled if $(1-\sum_{(l,t)\in k^*}\beta_{lt})>0$, we have $\beta_{lt}<1$ for all $(l,t) \in k^*$. Therefore, the number of packets transmitted on link $l$ at any time $t$ is at most $C_l$. Constraint \eqref{Schedule3} holds. 
By initialization and step \eqref{stepM:X}, constraint \eqref{Schedule4} holds.

When a new packet $m$ arrives, it will either be dropped or scheduled. If it is dropped, both $\Delta P$ and $\Delta D$ are 0. If it is scheduled, both \eqref{dual1} and \eqref{Schedule1} increase. With Lemma \ref{lemma:modified},the ratio between $\Delta P$ and $\Delta D$ is bounded by $2(1+\ln L) +\frac{B}{C_{min}}$. Therefore the competitive ratio of Algorithm \ref{Alg:ModifiedPD} is $(1, 2(1+\ln L) +\frac{B}{C_{min}})\rightarrow(1, 2(1+\ln L) )$, as $C_{min}\rightarrow\infty$.

\end{IEEEproof}

Thus, comparing with the result in \cite{multihoponline}, Algorithm \ref{Alg:ModifiedPD} achieves the optimal competitive ratio when $R=1$.

\section{A Fully Distributed Protocol for Implementation}
\label{section:distributed}
The two algorithms that we have proposed so far are both centralized algorithms. Specifically, when a packet arrives at a node, the node needs to have complete knowledge of all $\beta_{lt}$ of all links to find a valid schedule. Such information is usually infeasible to obtain. In this section, we propose a distributed protocol based on the design of Algorithm \ref{Alg:PD}. 

In our distributed protocol, the task of transmitting a packet to its destination is decomposed into two parts: First, when a packet arrives at a node, the node determines a suggested schedule based on statistics of past system history. This suggested schedule consists of the route for forwarding the packet, as well as a local deadline for each link. After determining the suggested schedule, the node simply forwards it to the first link of the route. On the other hand, when a link receives a packet along with a suggested schedule, the link tries to forward the packet to the next link in the suggested schedule before its local deadline. The link drops the packet when it cannot forward the packet on time.

To facilitate this protocol, each link keeps track of its own $\beta_{lt}$, which reflects the number of packets that are scheduled to be transmitted over link $l$ at time $t$. The value of $\beta_{lt}$ changes over time, as link $l$ schedules more and more packets to be transmitted at time $t$. Therefore, we define $\beta_{lt, \hat{t}}$ as the value of $\beta_{lt}$ when the current time is $\hat{t}$. Each link then measures $\gamma_{l,\tau}$ as the average of $\beta_{lt, t-\tau}$. In other words, when the current time is $t_0$, the expected value of $\beta_{lt}$ is $\gamma_{l,t-t_0}$. Link $l$ broadcasts its $\gamma_{l,\tau}$ periodically so that all nodes can estimate the values of $\beta_{lt}$.

We now describe how a node determines a suggested schedule upon the arrival of a packet. Suppose a packet arrives at time $t_0$. Following the Alg.~\ref{Alg:PD}, the node would like to find a valid schedule that maximizes $(1-\sum_{l,t:(l,t)\in k}\beta_{lt})$. In practice, the node does not know the exact value of $\beta_{lt}$. However, it knows that the expected value of $\beta_{lt}$ is $\gamma_{l,t-t_0}$. In our protocol, the node assumes that $\beta_{lt}=\gamma_{l,t-t_0}$, and then finds a valid schedule $k^*$ that maximizes $(1-\sum_{(l,t)\in k}\gamma_{l, t-t_0})$. Similar to Alg.~\ref{Alg:PD}, 
the node drops the packet if $(1-\sum_{(l,t)\in k^*}\gamma_{l, t-t_0})\leq 0$. If $(1-\sum_{(l,t)\in k^*}\gamma_{l, t-t_0})> 0$, then the node puts information of $k^*$ into the header of the packet, and forwards the packet to the first link in $k^*$. 
\begin{algorithm}
\caption{Distributed Implementation: Schedule Suggestion for Each Node}
\label{Alg:Distributed1}
\begin{algorithmic}[1]
  \FOR{each arriving packet $m$}
  	\STATE $t_0\leftarrow$ current time
 	\STATE $k^* \leftarrow \argmax_k {(1-\sum_{(l,t)\in k}\gamma_{l, t-t_0})}$ \label{stepd:k*}
  
	  \IF{$(1-\sum_{(l,t)\in k}\gamma_{l, t-t_0})>0 $} \label{stepd:checkalpha}
	 	 \STATE Put information of the suggested schedule $k^*$ in the header of packet $m$.
	 	 \STATE Forward the packet to the first link in $k^*$.
	  \ELSE
	  	\STATE Drop packet $m$.
	  \ENDIF
  \ENDFOR
\end{algorithmic}
\end{algorithm}

Since the actual value of $\beta_{lt}$ can be different from $\gamma_{l,t-t_0}$, there is no guarantee that a packet can be delivered on time using the valid schedule $k^*$ even if $(1-\sum_{(l,t)\in k^*}\gamma_{l, t-t_0})> 0$. Therefore, when a node determines a valid schedule $k^*$ for a packet, the valid schedule $k^*$ is treated only as a suggestion for links in $k^*$. Specifically, if $k^*$ contains an entry $(l^*, t^*)$, then the link $l^*$ interprets $k^*$ as a requirement that $l^*$ needs to forward the packet to the next link before $t^*$, or drops the packet. When $l^*$ obtains the packet, it still has the freedom to choose when to forward the packet, as long as the packet is forwarded before time $t^*$. 

Next, we discuss how each link determines the actual time to transmit each packet. Obviously, each link $l^*$ knows its own $\beta_{l^*t}$. From the design of Alg.~\ref{Alg:PD}, we can see that Alg.~\ref{Alg:PD} prefers to transmit packets when $\beta_{lt}$ is small. Our proposed policy is based on this principle. When a link $l^*$ receives a packet, it finds the entry $(l^*, t^*)$ from the valid schedule $k^*$ specified in the header of the packet. Link $l^*$ then finds a time $t_{tx}$ between the current time and $t^*$ that has the smallest $\beta_{l^*t}$, and transmits the packet at time $t_{tx}$. Alg.~\ref{Alg:Distributed2} describes the details of the policy for packet transmission.

\begin{algorithm}
\caption{Distributed Implementation: Packet Transmission for Each Link}
\label{Alg:Distributed2}
\begin{algorithmic}[1]
  \FOR {each packet $m$}  
    \STATE Upon $m$'s arrival at a link $l^*$, the link reads schedule information $k^*$ from the header of the packet. Let $t^*$ be the local deadline such that $(l^*,t^*)\in k^*$.
  	%\FOR{each link $l^*: (l^*,t^*)\in k^*$}
  		\STATE $ t^*_\text{tx} \leftarrow \argmin_{t_\text{tx}: t_\text{tx}\leq t^*} \beta_{l^*,t_\text{tx}}$
  		\IF{$\beta_{l^*,t^*_\text{tx}} <1$}
		  	\STATE $\beta_{l^*t^*_\text{tx}} \leftarrow \beta_{l^*,t^*_\text{tx}}(1+\cfrac{1}{C_{l^*}})+\cfrac{1}{(d_{l^*}-1)C_{l^*}}$

		  	\STATE Transmit packet $m$ on link $l^*$ at time $t^*_\text{tx}$.
		\ELSE
		  	\STATE Drop packet $m$.
		  	
	 	\ENDIF
	 %\ENDFOR
  \ENDFOR
  
%  \FOR{each link}
%  	\STATE Update $\gamma_{l\tau}$ as:
%  	$\gamma_{l\tau}=\frac{1}{T}\sum_{t=1}^{T} {\beta_{l,t+\tau,t}}$
  	
%  \ENDFOR

\end{algorithmic}
\end{algorithm}

\section{Simulation}
\label{section:simulation}
In this section, we evaluate the performance of our policies by simulation. We compare our algorithms with EDF policy and the policy, which we call Mao-Koksal-Shroff (MKS) online algorithm, proposed in \cite{multihoponline}. Both EDF policy and MKS online algorithm focus on packet scheduling, and are applicable only when the route of the packet is given. For these two policies, we assume that each packet is routed through the shortest path.

\begin{figure}[t]
	\begin{center}
	\includegraphics[width=3in]{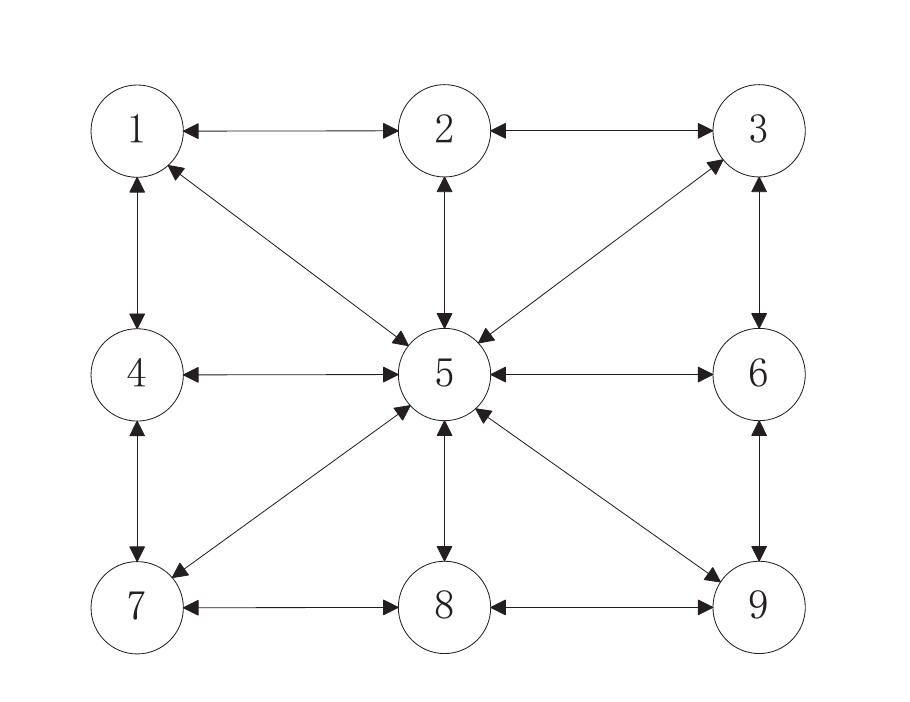}
	\caption{Network topology for a small network}
	\label{fig:simu1}
	\end{center}
\end{figure}

We first consider a small network as shown in Fig \ref{fig:simu1}. The network has 9 nodes from node 1 to node 9. There are directed arrows showing the directed links between nodes. All links have the same capacity $C=1$. We assume that there are 1000 packets arriving at the system. For each packet, the source node is chosen uniformly at random between node 1 to node 6, and the destination is chosen uniformly at random between node 7 to node 9. The inter-arrival time between packets are chosen to be 0 with probability 0.7 and 1 with probability 0.3. The deadline of each packet equals its arrival time plus a slack time. The slack time is chosen uniformly from integers between 2 and 6. %It is easy to see that in this network, any two nodes can be connected by as few as 1 links and as many as 6 links. Also, since we only specify source node and destination node, we use shortest path between source and destination for packet transmitting to analyze EDF policy and Shroff online algorithm. 

% The system used in \cite{multihoponline} deals with the network with homogeneous unit link capacity, thus here we also assume the network has unit link capacity and we use a similar system setting. Due to the complexity of simulating Shroff online algorithm, we use a simple system setting as follows. There are 1000 packets arriving at node 1 to node 6 with equal probability. Their destination are uniformly chosen from node 7 to node 9. The inter-arrival time between packets are chosen to be 0 with probability 0.7 and 1 with probability 0.3. The deadline of each packet equals its arrival time adding with a slack time. The slack time is chosen uniformly from integers between 2 and 6. It is easy to see that in this network, any two nodes can be connected by as few as 1 links and as many as 6 links. Also, since we only specify source node and destination node, we use shortest path between source and destination for packet transmitting to analyze EDF policy and Shroff online algorithm. 

%Under different link capacity $RC$, we calculate the ratio of successfully transmitted packets by our policies and compare our policy with two other policies. The detailed result is shown in Fig \ref{fig:result1}. 

\begin{figure}[t]
	\begin{center}
	\includegraphics[width=3.5in]{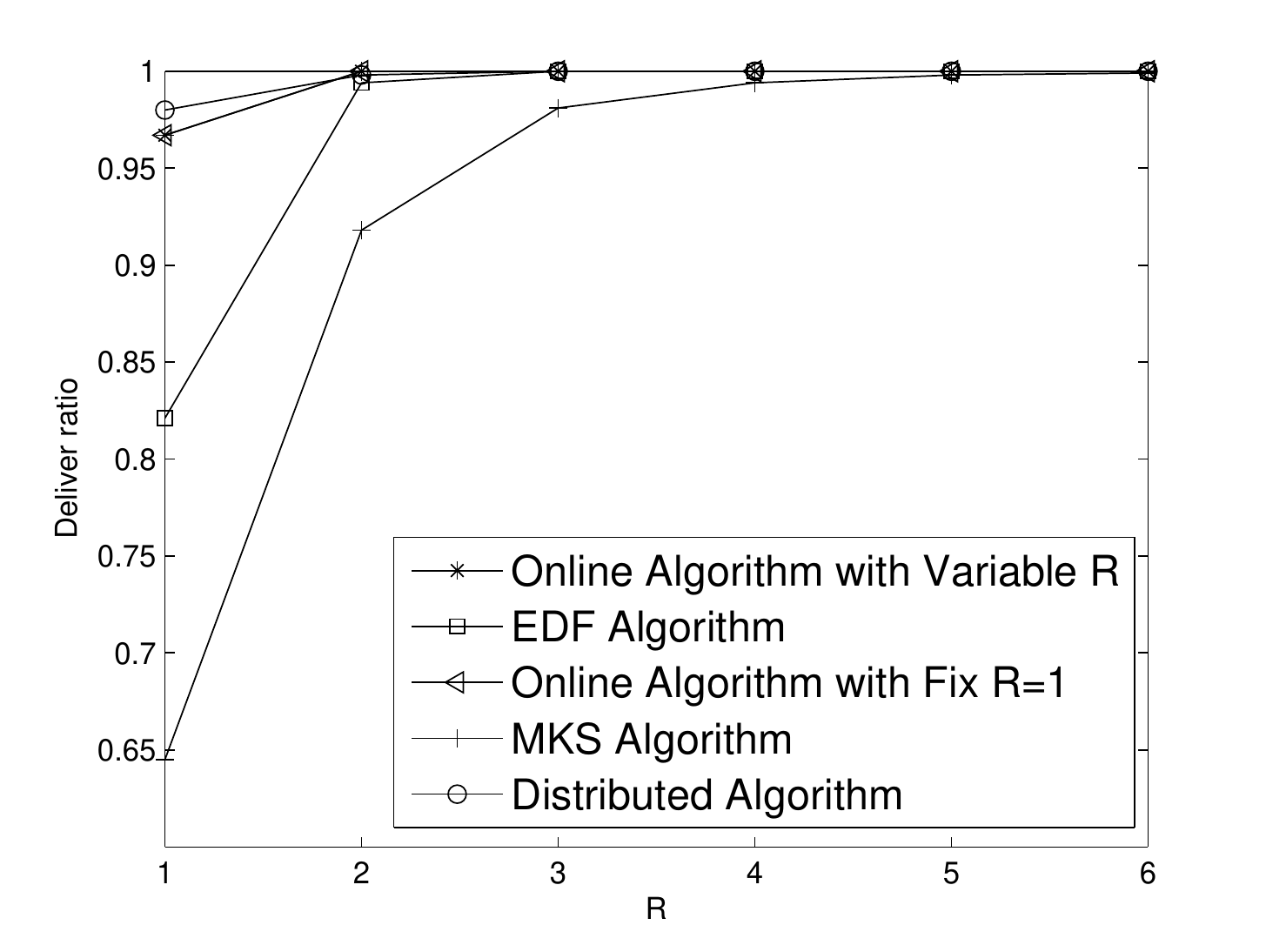}
	\caption{Deliver ratio comparison when all links have the same capacity.}
	\label{fig:result1}
	\end{center}
\end{figure}

Simulation results for different values of $R$ are shown in Fig. \ref{fig:result1}.
%The result shows the packet deliver ratio of the four policies with different link capacity $RC$. 
 From the result, we can see that all our three policies outperform two other current policies. %This is because using shortest path could cause more congestions on the network, while our policies are able to choose routes wisely based on current traffic on all links. 
 From the figure, we can see that all our policies are able to deliver all  packets when $R$ is 2. On the other hand, EDF is able to deliver all packets when $R=3$, and MKS can deliver all packets only when $R$ is as large as 6. We also note that both EDF and MKS are centralized policies. The fact that our distributed algorithm performs better than these two centralized policies further highlights the superiority of our algorithms.
 %Also, in this network setting, the distributed algorihtm has better performance than distributed algorithms. In the result, our algorithms with variable $R$ and fixed $R$ have identical result. This is because the network has unit link capacity. Thus the link load $\frac{n}{C_l}$ is either $0\%$ or $100\%$. We are not able to increase $\beta_{lt}$ slower as we stated in Section \ref{section:modified alg}. Thus the two algorithms have the same performance. When the capacity increases to 2, both policies can deliver all packets.

\begin{figure}[t]
	\begin{center}
	\includegraphics[width=3.5in]{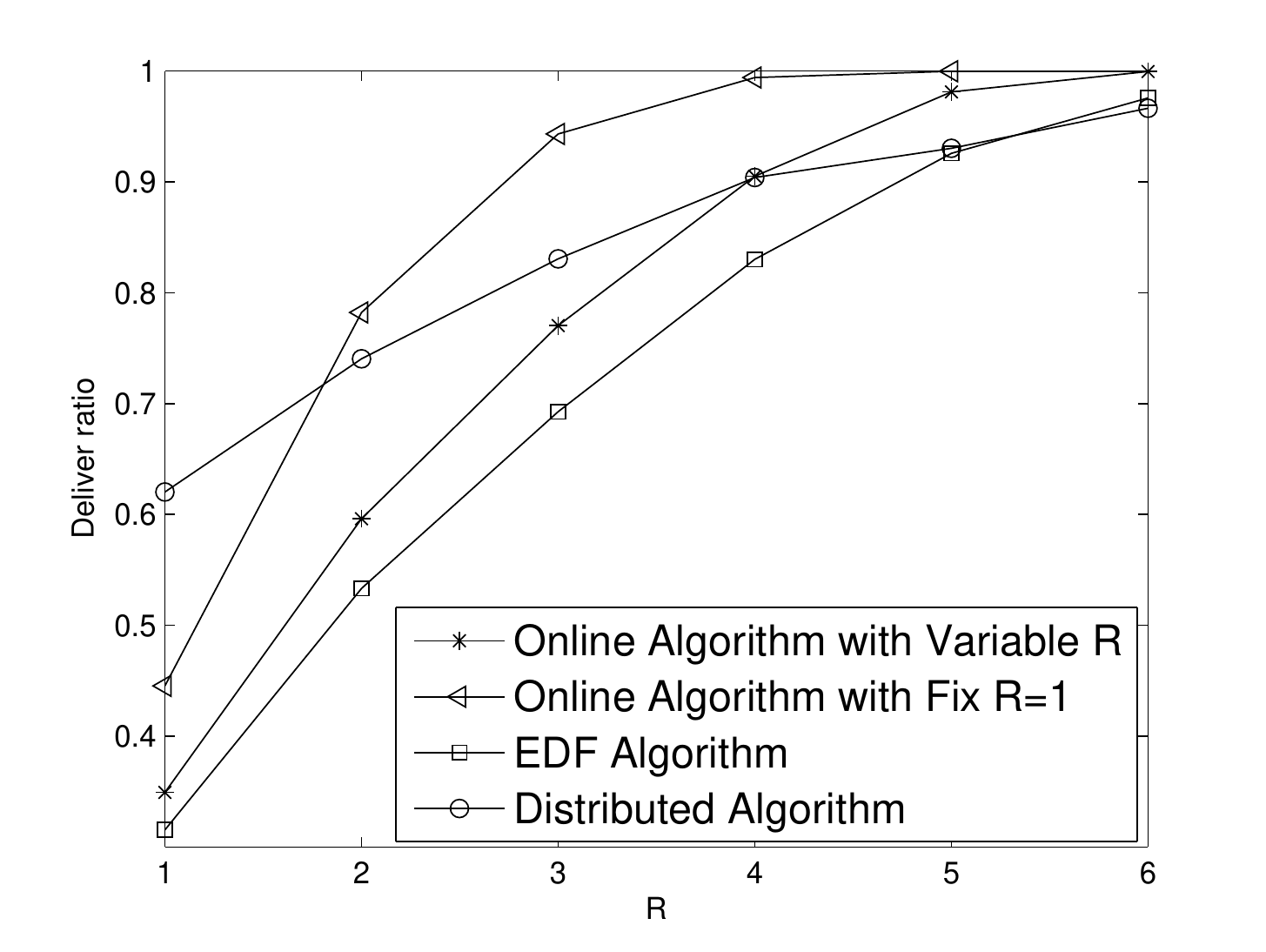}
	\caption{Deliver ratio comparison when different links have different capacities.}
	\label{fig:result2}
	\end{center}
\end{figure}

Next, we consider that different links can have different capacities. Since MKS requires all links to have the same capacity, we only compare our policies against EDF. The network topology is also shown in Fig \ref{fig:simu1}. We assume that, when $R=1$, the link capacity are integers uniformly chosen from 5 to 10. There are 10000 i.i.d packets to be delivered. At the beginning of each time slot, there are a certain number of packets arriving the system. The source node and destination node are both chosen from node 1 to node 9 with equal probability and destination node is not allowed to be the same with source node. The number of packets is randomly chosen between 100 and 500. Each packet has a slack time between arrival and deadline, which is uniformly chosen from $[2,6]$. The result is shown in Fig \ref{fig:result2}. Once again, we see that our policies, including the distributed algorithm, perform much better than EDF in most cases.

%From the result, we can see that as the origin and destination node increase, the routing gets more complicated, and number of packets increases, the congestion on the network increases. By comparing our proposed two centralized policies, we can verify that the order optimal policy (online algorithm with fix $R=1$) has better performance than the other. In this setting, both our policies outperform EDF policy. Also, the distributed policy provide a satisfied delivery ratio. With original link capacity, the distributed algorithm even outperform all three other policies.

%\begin{figure}[t]
%	\begin{center}
%	\includegraphics[width=3.5in]{figure/simu2.eps}
%	\caption{10 $\times$ 10 grid network.}
%	\label{fig:simu2}
%	\end{center}
%\end{figure}

%Finally, we simulate and compare the policies in a 10 $\times$ 10 grid network as shown in Fig. \ref{fig:simu2}. The link capacity are integers uniformly chosen from 1 to 5 when $R=1$. There are $10^5$ i.i.d packets to be delivered. At the beginning of each time slot, the arriving number of packet is randomly chosen between 100 and 500. The source node and destination node are both chosen from node 1 to node 100 with equal probability. Each packet has a slack time uniformly chosen from $[5,20]$. In this large grid network, we also compare our policies with EDF. The result is shown in Fig \ref{fig:result3}. We see that our centralized policies perform much better than EDF. 

%\begin{figure}[t]
%	\begin{center}
%	\includegraphics[width=3.5in]{figure/result4.eps}
%	\caption{Deliver ratio comparison}
%	\label{fig:result3}
%	\end{center}
%\end{figure}

\section{Conclusion}
\label{section:conclusion}
In this paper, we study the multi-hop network scheduling problem with end-to-end deadline and hard transmission rate requirement. Given the capacity of each link in the network, we aim to find out how much capacity we need to increase to guarantee the required ratio of packets can be successfully transmitted to its destination before its deadline without knowing the packet arrival sequences in advance.

We have proposed an online algorithm which works for both fix route and non-fix route network. The algorithm is proved to be $(R,1+\frac{L}{e^R-1} )$-competitive, where $L$ is the length of the longest path. We have also showed that the complexity of our algorithm is $O(ET)$, where $E$ is the total number of links and $T$ is the largest slack time. 
Next, we have showed that any online algorithm cannot be better than $(R, 1+\frac{L-2e^R}{(L+1)e^R-L})$-competitive. When both $L$ and required deliver rate are large, our policy requires at most twice as much capacity as the lower bound. 
In addition, We have proposed an online algorithm for fixed capacity network. When the capacity cannot be increased, our algorithm is proved to be $(1, O(\log L))$-competitive, which is also an order-optimal policy. For practical implementation of our centralized algorithm, we have proposed a heristic for distributed algorithm so that each node can make decisions without requiring real-time information from all other nodes. In addition to the theoretical results, we compare our policies with two other online policies, including the widely-used EDF policy and a recent proposed policy, by simulation. The results show that the performance of our policies are better than the other two policies. Also the result shows that the distributed algorithm still provide a good delivery ratio.

\section{Acknowledgment}
This material is based upon work supported in part by the U. S. Army Research Laboratory and the U. S. Army Research Office under contract/grant number W911NF-15-1-0279 and NPRP Grant 8-1531-2-651 of Qatar National Research Fund (a member of Qatar Foundation).
 
%e extend our network model from the previous studies and our model has several significant differences compared to the network in those research. First, we do not assume any knowledge of the future arrival packet, which better resemble the practical cases.  Second, our network is not limited to fix route system. Once we obtain the deadline and destination node information, our algorithm can find a route for the packet and schedule the time that the packet is transmitted through each link on this route. Third, instead of simply using unit capacity links, we further consider that different links may have different capacity. Therefore, our model is more realistic compared to previous ones. 

\balance

%\bibliography{reference}
%\bibliographystyle{ieeetr}
\end{document}